\documentclass{PoS}
\title{Fisher zeros and conformality in lattice models}

\ShortTitle{Fisher zeros and conformality in lattice models}

\author{\speaker{Yannick Meurice}$^a$\footnote{Current email address: yannick-meurice@uiowa.edu}, Alexei Bazavov $^b$, Bernd A. Berg $^c$, Daping Du $^{a,d,e}$, Alan Denbleyker $^a$, Yuzhi Liu $^{a,d}$, Donald K. Sinclair $^{a,f}$, Judah Unmuth-Yockey$^a$, Haiyuan Zou $^a$\\
\llap{$^a$} Department of Physics and Astronomy, University of Iowa, Iowa City, IA 52240, USA\\
\llap{$^b$} Brookhaven National Laboratory, Upton, NY 11973, USA\\
\llap{$^c$}Department of Physics, Florida State University\, Tallahassee, FL 32306, USA\\
\llap{$^d$} Fermi National Accelerator Laboratory, Batavia, IL 60510, USA \\
\llap{$^e$} Physics Department, University of Illinois, Urbana, IL 61801, USA\\
\llap{$^f$}HEP Division, Argonne National Laboratory, Argonne, IL 60439, USA\\

}

\abstract{Fisher zeros are the zeros of the partition function in the complex $\beta=2N_c/g^2$  
plane. When they pinch the real axis, finite size scaling allows one to distinguish 
between first and second order transition and to estimate exponents. On the other hand, 
a gap signals confinement and the method can be used to explore the boundary of the conformal window. 
We present recent numerical results for 2D $O(N)$ sigma models, 4D $U(1)$ and 
$SU(2)$ pure gauge and $SU(3)$ gauge theory with $N_f=4$  and 12 flavors. We discuss attempts to 
understand some of these results using analytical methods. We discuss the 2-lattice 
matching and qualitative aspects of the renormalization group (RG) flows in the Migdal-Kadanoff approximation, in particular how RG flows starting at large $\beta$ seem to move around regions where bulk transitions occur. We consider the effects of the boundary conditions on the nonperturbative part of the average energy and on the Fisher zeros for the 1D $O(2)$ model.}

\FullConference{The 30th International Symposium on Lattice Field Theory\\
                 June 24-29,  2012\\
                 Cairns, Australia}
\begin{document}

\section{Content of the talk}

With the ongoing effort at the LHC, there has been  a renewed interest in the phase diagram of lattice gauge
theory models. 
The location of the conformal windows
for several families of models have triggered intense discussions.
Different numerical and analytical techniques have been applied
to QCD-like models with a large number of fermion flavors
\cite{one,appelquist09, hasenfratz09,fodor09, Deuzeman:2009mh,Hasenfratz:2010fi,Fodor:2011tu}
or with fermions in higher representations
\cite{shamir08,2010PhRvD..81k4507K,DeGrand:2011qd,min}. 
See also \cite{DeGrand:2010ba,Ogilvie:2010vx,Sannino:2009za,tech} for recent reviews of results and expectations.
It is important to understand the critical behavior of lattice models from various points of view. 
It was  proposed to consider complex extensions \cite{Denbleyker:2010sv,PhysRevD.83.056009,Liu:2011zzh} of the picture of confinement proposed by Tomboulis \cite{Tomboulis:2009zz}. It was observed that 
the Fisher's zeros, the zeros of the partition function in the complex $\beta$ plane,  
determine the global properties of the 
complex RG flows. In the case where a phase transition is present, the scaling properties of the zeros \cite{alves90b,Jersak:1996mn,Jersak:1996mj,janke04} allow us 
to distinguish between a first and second order phase transition. In the following, we briefly review the Finite Size Scaling (FSS) of the 
 Fisher's zeros. We then discuss numerical results for 
$U(1)$ in 3 and 4D,
$SU(2)$ in 4D with $\beta _{Adjoint}$, 
 $SU(3)$ in 4D $N_f=4$ and 12 flavors. We also discuss the 
 effect of boundary conditions in the $O(2)$ sigma model and recent effort to relate these questions to perturbative expansions. 

\section{Fisher's zeros and Finite Size Scaling (FSS)}
Fisher's zeros provide information about FSS. The basic principle is the 
decomposition of the partition function \cite{niemeijer76} at finite volume $L^D$ into a singular and a regular part:
\begin{eqnarray} Z&=&Z_{sing.} {\rm e}^{G_{bounded}} \\
Z_{sing.}&=&{\rm e}^{-L^D f_{sing.}}
\end{eqnarray}
Under a 
RG transformation, the lattice spacing $a$ increases by a scale factor $b$, and 
\begin{eqnarray}
L &\rightarrow& L/b \\
f_{sing.}&\rightarrow& b^D f_{sing.}\\
Z_{sing.}&\rightarrow& Z_{sing.} \ .
\end{eqnarray}
This has the important consequence \cite{Itzykson:1983gb} that
the zeros of the partition functions are RG invariant.
The nonlinear scaling variables (e. g. $u=\beta -\beta_c$, $\dots $) 
transform homogeneously: 
$u_i\rightarrow \lambda_i u_i$. 
For the 
relevant variables, we use the notation $\lambda_i = b^{1/\nu_i}$, and for the irrelevant variables $\lambda_j = b^{-\omega _j}$. 
The RG invariance of $Z_{sing.}$ means that 
\begin{equation}
Z_{sing.}=Q(\{u_iL^{1/\nu_i}\},\{ u_jL^{-\omega _j}\})
\end{equation}

For a single relevant variable $u\simeq \beta -\beta_c $,  we have $Z_{sing}=Q(uL^{1/\nu})$.
The complex equation $Z=0$ can be written as two real equations for two real variables and generic solutions are 
isolated points. 
\begin{equation}
Z=0 \Rightarrow uL^{1/\nu}=w_r \  {\rm with} \ r=1, 2, \dots
\end{equation}
This implies the approximate form for the zeros: 
\begin{equation}\beta_r(L)\simeq \beta_c +w_rL^{-1/\nu}\end{equation}
There are many examples, where these discrete solutions follow approximate lines or lay inside cusps. In the infinite volume limit, 
the set of zeros may (or may not) separate the complex plane into two or more regions. For a first order transition $\nu$ is replaced by $1/D$. 
For confining models where there is no phase transition on the real coupling axis between weak and strong coupling, it is sometimes possible to 
construct numerically RG flows in the complex $\beta$ plane \cite{Denbleyker:2010sv,PhysRevD.83.056009,Liu:2011zzh}. These examples show that the 
Fisher's zeros can act as ``gates" for the complex RG flows and govern their global behavior. Since it is much easier to calculate the Fisher's zeros than 
the complex RG flows we shall now describe model calculations of the zeros. 
\section{Model calculations}

\begin{figure}\begin{center}
\includegraphics[width= 2.6in]{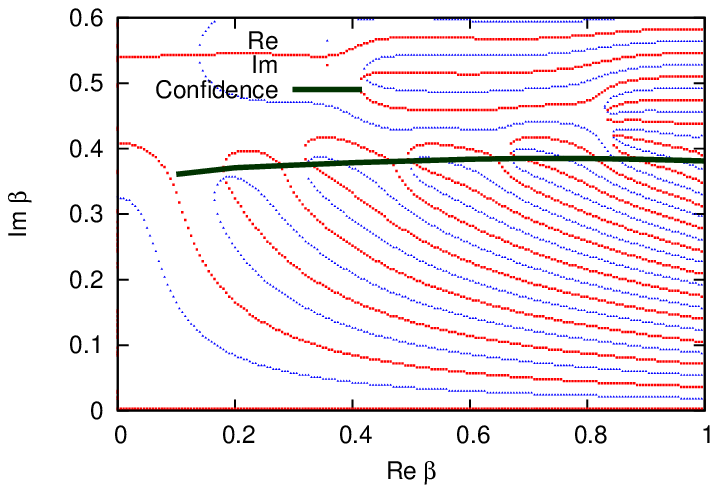}
\includegraphics[width= 2.6in]{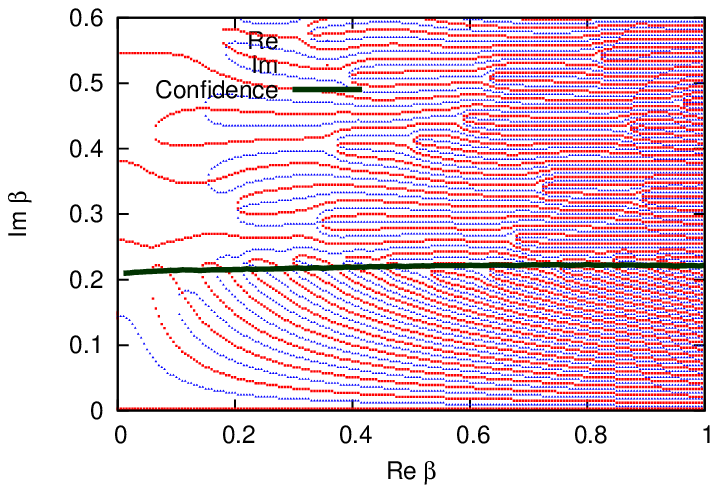}
\caption{\label{fig:3u1}Fisher's zeros for $U(1)$ on $L^3$ lattices ($L$=4 (left) and  6 (right)).  The zeros of the real (imaginary) part are represented by the blue (red) curves and the region of confidence is below the green line (zeros near or above this line are not reliable).} 
\end{center}
 \end{figure}

We showed previously that for 2D $O(N)$ models in the large-$N$ limit, 
the RG flows go directly from weak coupling to strong coupling \cite{Denbleyker:2010sv,PhysRevD.83.056009}. 
This means that the theory has a mass gap. 
We would expect the same result  for 
3D $U(1)$. Consistently, using canonical simulations and histogram reweighting, we found no zeros near the real axis. This is illustrated in 
Fig. \ref{fig:3u1} where anything above the green line is considered nonreliable.

The case of the 4D $U(1)$ is more delicate. There is a transition but is it first or second order? 
Despite some earlier suggestions of a second order transition \cite{Jersak:1996mn,Jersak:1996mj}, there seems to be a general agreement that the 
transition is first order \cite{Campos:1998jp,Arnold:2002jk,Vettorazzo:2003fg}. 
This question has been revisited recently from the point of view of Fisher zeros \cite{PhysRevD.85.056010}. 
For $L^4$ lattices, the average plaquette distribution has a double peak distribution with equal heights at a pseudo-critical $\beta_S$. 
A double peak does not always mean a first order transition. 
We consider a simple example where the plaquette distribution at $\beta_S$ is a superposition of two Gaussians:
\begin{equation}
n(S){\rm e}^{-\beta_S S}\propto ({\rm e}^{-(1/2\sigma^2)(S-S_1)^2}+{\rm e}^{-(1/2\sigma^2)(S-S_2)^2})
\end{equation}
The zeros are located at $\beta_r =\beta_S+i2\pi(2r+1)/(S_2-S_1)$. 
If $(S_2-S_1)\propto L^D$, we have a first order phase transition (latent heat) and Im$\beta_1 \propto L^{-D}$.
However, if $(S_2-S_1)\propto L^{D-\zeta}$, then the width of the double peak in the {\it average} plaquette goes to zero at infinite volume and Im$\beta_1 \propto L^{-1/\nu}$ with 
$\nu=1/(D-\zeta)$.
For the actual $U(1)$ case for small $L$, the distance between the peaks slowly decreases with increasing volume. 
In the infinite volume limit, the width of the double peak distribution of the average plaquette goes to a nonzero limit (latent heat) for a first order phase transition
and  to zero as an inverse power of $L$ for a second order transition. At this point, better statistics at large volumes are necessary in our numerical construction of the density of states to discriminate between the two scenarios, however fits of the imaginary part of the lowest zero described in Ref. \cite{PhysRevD.85.056010} show
a goodness of fit $Q$=0.43  for $L^{-4}$ leading behavior and  $Q<10^{-8}$ for $L^{-3.08}$ which favors the first order scenario. 

\begin{figure} \begin{center}
\includegraphics[width= 2.8in]{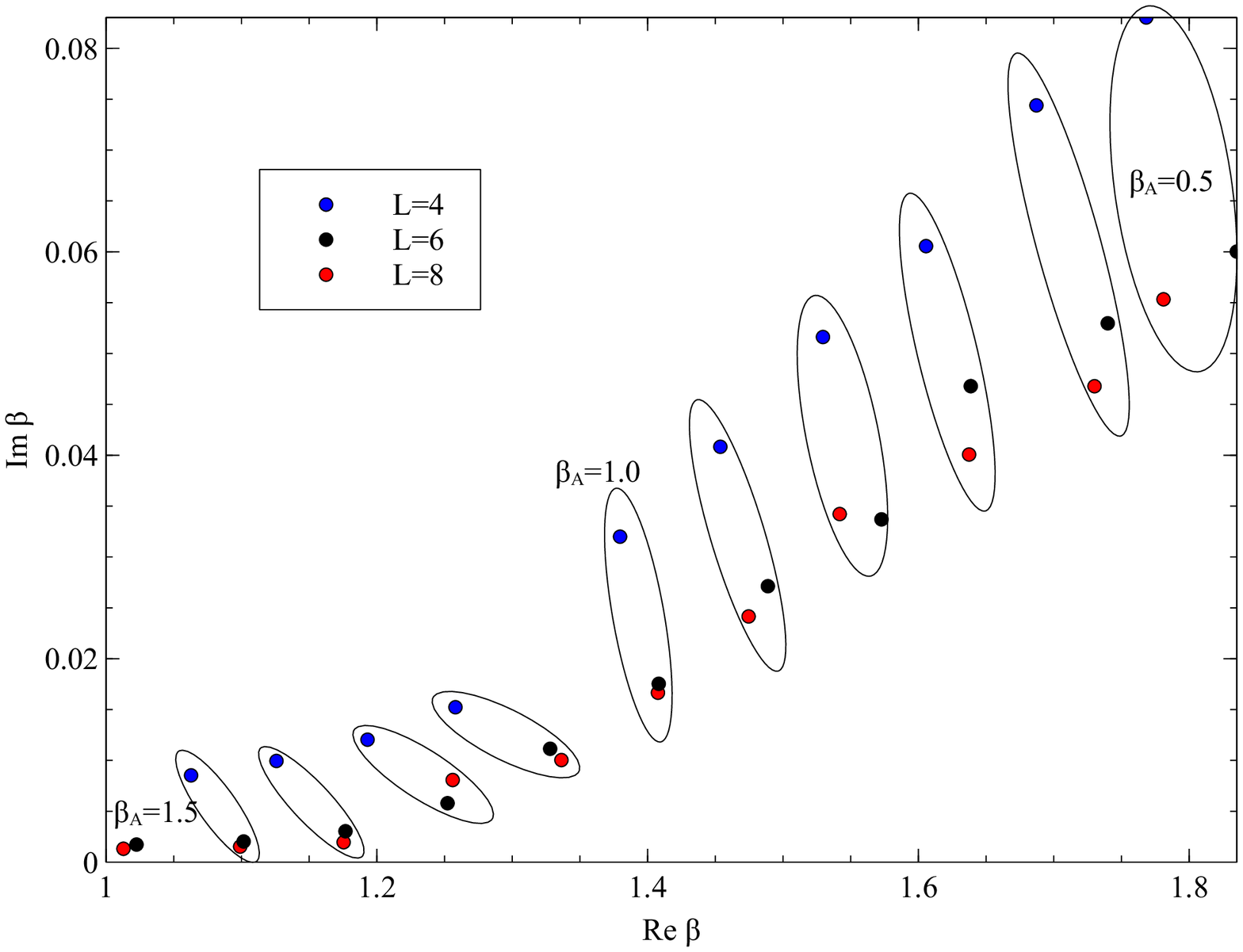}
\caption{\label{fig:lowest}Lowest zeros for $\beta_{Adjoint}$= 0.5, 0.6, ..., 1.5 for $L^4$ lattices with $L$= 4,6,and 8. } \end{center}
 \end{figure}

For pure gauge $SU(2)$, it is possible to create double peak distributions by adding an adjoint term with a coefficient $\beta_{Adjoint}$ sufficiently large and positive 
to the Wilson action. As a positive $\beta_{Adjoint}$ is increased, the Fisher's zeros go down as illustrated in Fig. \ref{fig:lowest}. There is a clear change of behavior for $1.0<\beta_{Adjoint}<1.1$. Subleading effects are important and larger volumes are being studied in order to draw conclusions regarding the scaling of the real and imaginary part of the lowest zeros. 
 
\begin{figure} \begin{center}
\includegraphics[width= 2.8in]{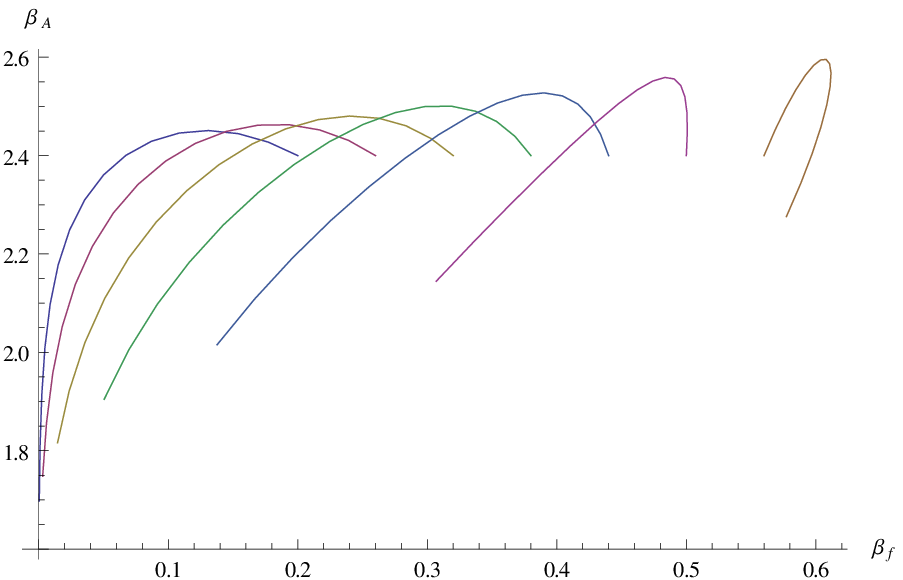}
\includegraphics[width= 2.8in]{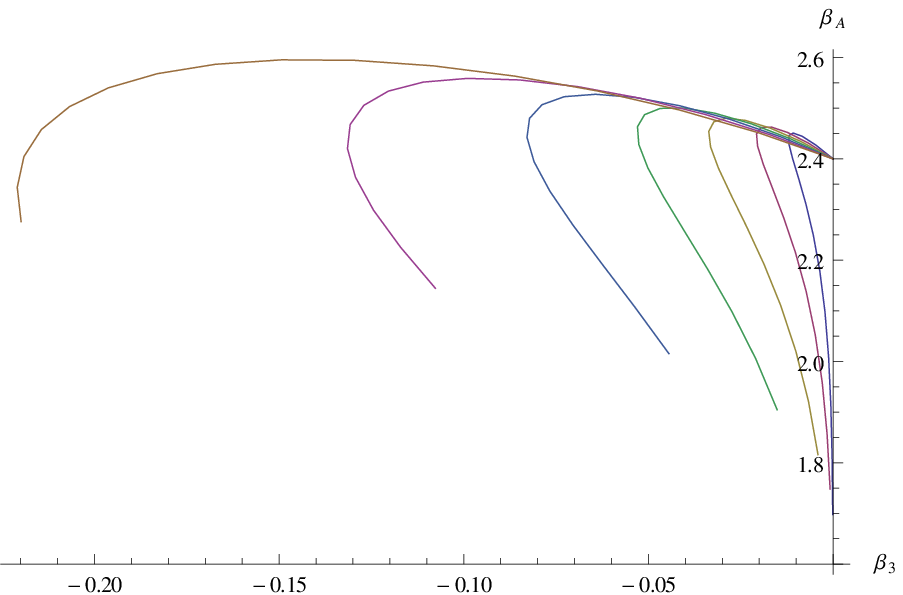}
\caption{\label{fig:mk} Projection of the MK flows in the ($\beta_{1/2},\beta_{1}$) and ($\beta_{3/2},\beta_{1}$) planes. } 
\end{center}
 \end{figure}

 We have started investigating the possibility of doing 2-lattice matching for $SU(2)$ with a mixed action \cite{alan}.  
 A first step consists in considering the effect of $\beta_{Adjoint}$ on RG flows starting at weak coupling. 
 This question can be addressed using the Migdal-Kadanoff (MK) approximation. 
 Using expansions with 20 characters, we found that the RG flows seem to go around possible boundaries in order to reach the  strong coupling fixed point (all $\beta$'s = 0). 
 Projections in two planes are shown in Fig. \ref{fig:mk} and are in qualitative agreement with this idea and flows described by \cite{bitar82}.
 
 It is also possible to test the accuracy of the MK approximation by checking the 
 2-lattice matching. If we had an exact RG transformation between a fine $(2L)^4$ lattice and a coarse $L^4$ lattice, then 
the $2R\times 2R$ Wilson loops for the $(2L)^4$ lattice and the $R\times R$ Wilson loop on a $L^D$ lattice with effective couplings obtained by 
the RG transformation should match exactly. However, if we use the MK approximation, the matching is not very accurate especially at large coupling as shown in the Table \ref{tab:adj}. 
This could also be used as a test for the  Cheng-Tomboulis improvement proposed at this conference \cite{ct} . 

\begin{table}[h!]
\begin{center}
  {\tiny \begin{tabular} {| l | c | c | c | c | c | c | c | c |}
    \hline
    Volume  & $b$ & $\beta_F$ & $\beta_A$   & $\beta_{3/2}$ & $\beta_2$     & $P_{size}$ & $\langle P \rangle$ & $\sigma$ \\ \hline
		$8^4$ &           &  2.40000   & 0.00000     &               &               & $2x2$      & 0.7766      & 0.00672  \\
	$4^4$ & $2$       &  0.955274  &-0.0496152   & 0.003759328   &-0.000310275   & 1x1           & 0.7710      & 0.01226  \\ \hline

	$8^4$ &           &  2.40000   & 0.00000     &               &               & $4x4$      & 0.9009     & 0.09007  \\
	$4^4$ & $2$       &  0.955274  &-0.0496152   & 0.003759328   &-0.000310275   & $2x2$      & 0.9973     & 0.01283  \\ \hline
	$8^4$ &           &  4.80000   & 0.00000     &               &               & $2x2$      & 0.4016      & 0.00369  \\
		$4^4$ & $2$       &  4.47578   &-0.728286    & 0.188086      & 0.055336      &    1x1        & 0.2225      & 0.00655  \\ \hline

	$8^4$ &           &  4.80000   & 0.00000     &               &               & $4x4$      & 0.5670      & 0.12841  \\
		$4^4$ & $2$       &  4.47578   &-0.728286    & 0.188086      & 0.055336      & $2x2$      & 0.5144      & 0.01799  \\ \hline
 \end{tabular}}   \end{center}
 \caption{Wilson loops on fine and coarse lattices.}
\label{tab:adj}
 \end{table}
 
\begin{figure}
\includegraphics[width=0.45\textwidth]{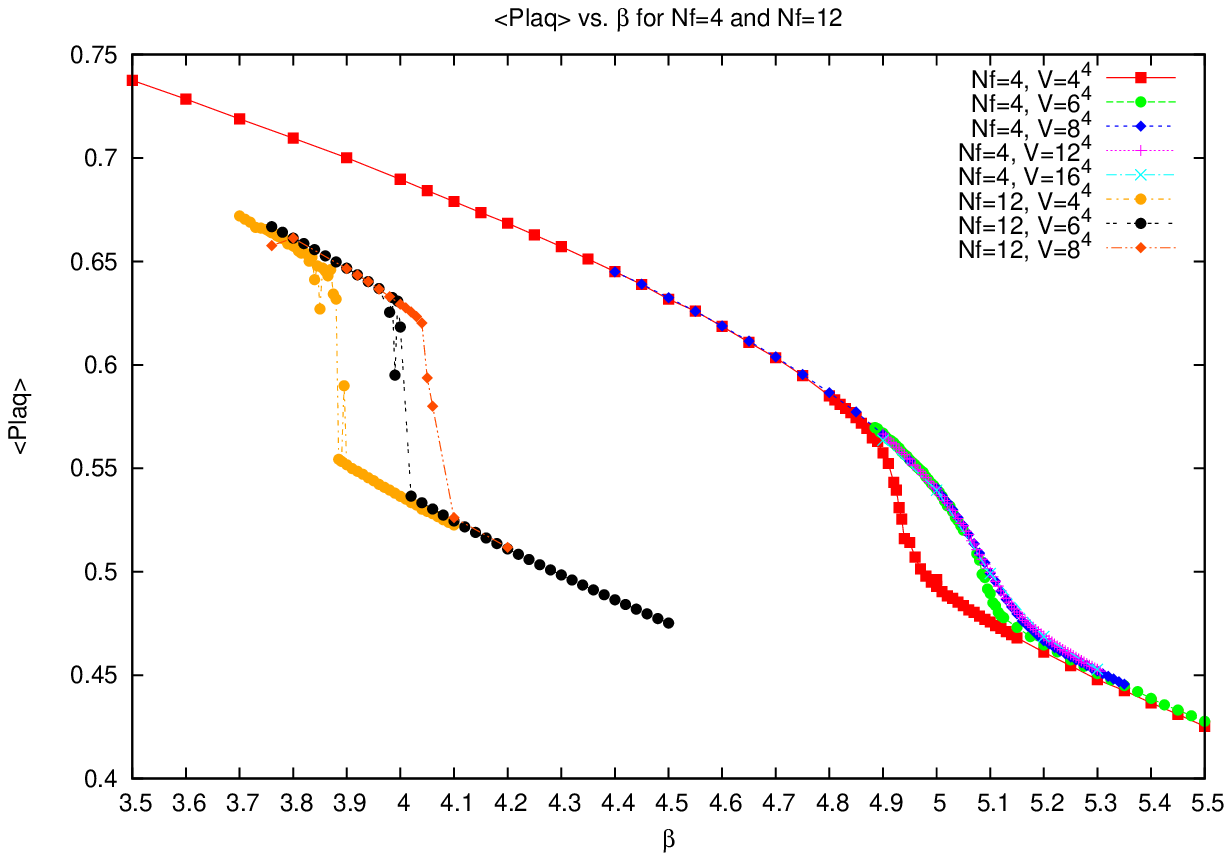}
\includegraphics[width=0.45\textwidth]{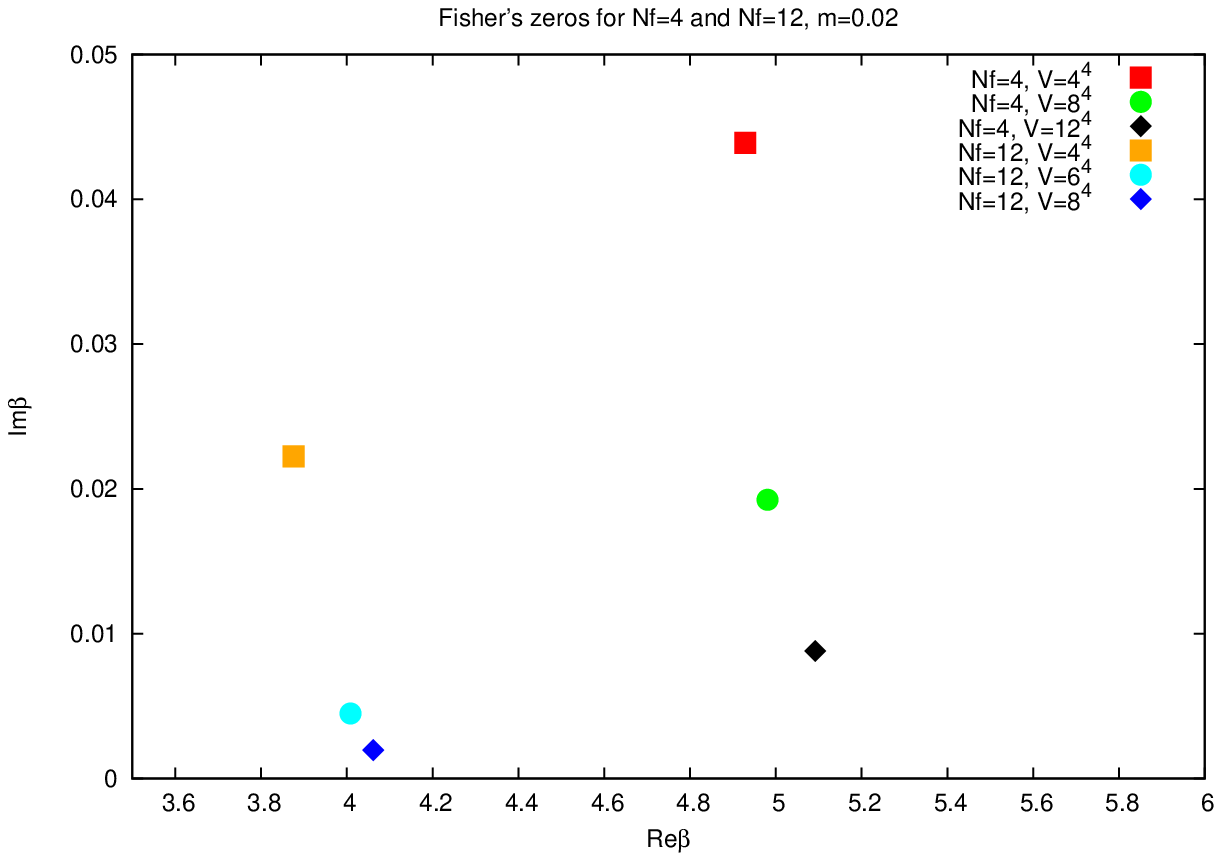}
\caption{Average plaquette and Fisher's zeros for $SU(3)$ with $N_f$= 4 and 12 at different volumes.}
\label{fig:12} 
 \end{figure}

The Fisher's zeros for $SU(3)$ with $N_f$= 4 and 12 quarks in the fundamental representation have been investigated \cite{yl}.
 We used the standard Wilson action and unimproved staggered fermion action with the Rational Hybrid Monte Carlo (RHMC) algorithm.  
 We started with relatively small symmetric lattices and up to 50,000 configurations. 
 The bare quark mass is set to be $m_q=0.02$ for now. For $N_f=12$, we found a discontinuity 
 for the plaquette near $\beta\simeq 4.1$ for $V = 8^4$ lattices as shown in Fig. \ref{fig:12} left panel. The plaquette histories showed the characteristic hysteresis behavior. For comparison, we show the crossover for $N_f=4$. 
The lowest zeros are shown in Fig. \ref{fig:12} right panel. Possible hypothesis to be tested are that for $N_f=12$, the imaginary part scales like $L^{-4}$, which signals a first order phase transition
 and that the real part increases like $\log (L)$. Nonlinear effects seem important at small volume and larger volume calculations are in progress. As already noticed 
 in Ref. \cite{one}, the chiral condensate has a discontinuity near the same value of $\beta$ as the plaquette. More recently, it was shown that improved actions can create a second discontinuity with a 
broken single-site shift symmetry between the two transitions \cite{anna,pal}. 

\begin{figure}
 \begin{center}
	\includegraphics[width=3in]{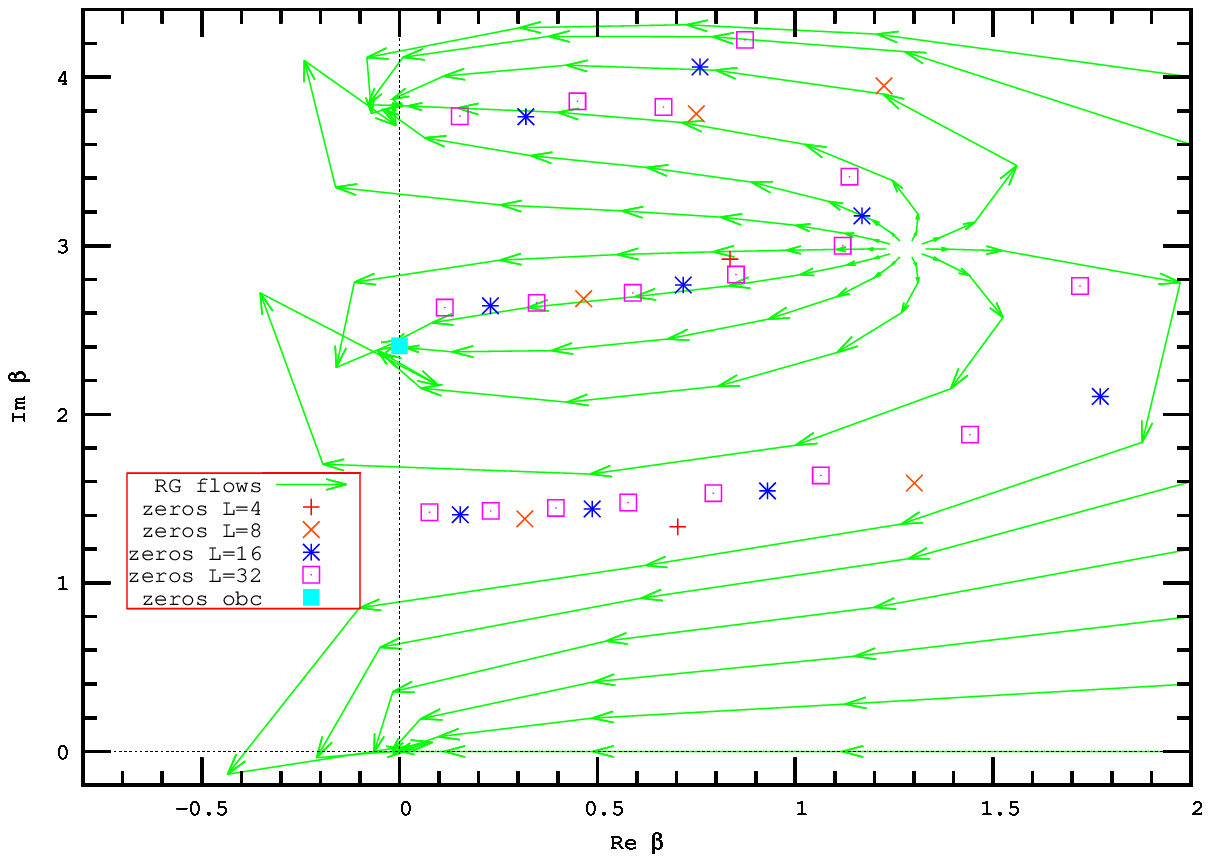}
\end{center}
		\caption{MK complex flows for the system (o.b.c) and zeros from the two different boundary conditions }
		\label{fig:o2}
	\end{figure}

We also studied the Fisher's zeros for the 
 1D O(2) with $L=4, 8, 16, 32$ \cite{hz}.  As shown in Fig. \ref{fig:o2}, the zeros are very different for open (o.b.c) and periodic boundary conditions (p.b.c). 
The MK complex flows can also be constructed and are also shown in Fig. \ref{fig:o2}. 
At finite volume, the nonperturbative parts of the average energy are very different for open and periodic boundary conditions. 
It was found that $|(E-E_{PT})/E| \propto {\rm e}^{-2\beta}$  for open boundary condition and $\propto  {\rm e}^{-\beta E_v}$ for periodic boundary condition,
where $E_v$ is the energy of the periodic solution of the classical equation of motion with winding number 1. 
Hadamard series have been constructed to improve the accuracy at strong coupling. 
	
\section{Conclusions}

Much progress has been made in finding reliable ways to locate the Fisher's  zeros of various models. 
A consistent picture of confinement in terms of complex RG flow is emerging. 
Much work remains to be done for understanding multiflavor models within this approach. 
Better analytical approaches (based on improved RG or weak coupling expansions) are needed. 
The FSS of zeros is simple, however subleading corrections are important (at least for unimproved actions). 
We plan to monitor the effects of improvement on the zeros by turning on improvement adiabatically.

This 
research was supported in part  by the Department of Energy
under Contracts No. FG02-91ER40664, DE-FG02-97ER41022, DE-FG02-12ER41871 and DE-AC02-98CH10886.

\end{document}